\documentclass{aa}  

\usepackage{graphicx}
\usepackage{txfonts}
\usepackage{lipsum}
\usepackage{subcaption}         
                               
\usepackage{lscape}             
                                
\usepackage{placeins}          

\usepackage[pdftex]{hyperref}

\hypersetup{
    colorlinks=true,       
    linkcolor=blue,      
    citecolor=blue,      
}

\begin{document}

\title{Modelling solar radial velocities and photometric variability with \texttt{SOAP}}

\author{A. Barka\inst{1,2}\thanks{Email: \href{mailto:alba.barka@astro.up.pt}{alba.barka@astro.up.pt}}
      \and E. Cristo\inst{1}
      \and \^A. R. G. Santos\inst{1,2,3}
      \and N. C. Santos\inst{1,2}
      \and K. Al Moulla\inst{1}\thanks{SNSF Postdoctoral Fellow}
      \and T. Barata\inst{4,5}
      \and R. Gafeira\inst{4,6}
      \and M. Cretignier\inst{7}
}

\institute{Instituto de Astrofísica e Ci\^encias do Espaço, Universidade do Porto, CAUP, Rua das Estrelas, 4150-762 Porto, Portugal
    \and Departamento de Física e Astronomia, Faculdade de Ci\^encias, Universidade do Porto, Rua do Campo Alegre, 4169-007 Porto, Portugal 
    \and Universit\'e Paris-Saclay, Universit\'e Paris Cit\'e, CEA, CNRS, AIM, 91191, Gif-sur-Yvette, France
    \and Instituto de Astrofísica e Ciências do Espaço, Department of Physics, University of Coimbra, Rua Larga, PT3040-004 Coimbra, Portugal
    \and Departamento de Ci\^encias da Terra, FCTUC, Universidade de Coimbra, 3004-531 Coimbra, Portugal
    \and Geophysical and Astronomical Observatory, Faculty of Science and Technology, University of Coimbra, Rua do Observatório s/n, PT3040-004, Coimbra, Portugal
    \and Department of Physics, University of Oxford, OX13RH Oxford, UK
}

\date{}

  \abstract
   {Stellar activity remains one of the main limitations in the detection of Earth-like planets using radial velocity (RV) measurements. The Sun, as the only star for which surface features can be spatially resolved, offers a unique testbed for studying the impact of active regions on RV and photometric variability.}
   {Using \texttt{SOAPv4} (Spot Oscillation And Planet), we modelled solar RV and photometric variability induced by spots and faculae over long timescales. Our goal is to verify whether present-day, state-of-the-art models of the cross-correlation function correctly reproduce the observed variability. Moreover, we aim to assess how the choice of input data and identification technique influences the agreement between simulated and observed signals.}
   {To simulate solar RV and photometric time series, we first identified active regions in SDO images. This was done using mathematical morphological transforms applied to SDO/HMI and AIA images. Mathematical morphological identification was validated against other state-of-the-art identification methods. Using these inputs, we ran \texttt{SOAPv4} to simulate solar RVs and photometry, and we validated the results with HARPS-N RV observations, as well as with VIRGO/SPM photometric measurements.}
   {The simulations that use mathematical morphological identification achieved the best match with the observed RV time series, yielding residuals with a measured standard deviation of $\sim0.91$ m/s. Other state-of-the-art methods produced higher filling factors and, consequently, larger discrepancies. The photometric simulations reproduced the overall variability trends.}
   {We demonstrate that mathematical morphological transforms accurately identify solar active regions. Using these inputs, \texttt{SOAPv4} reproduces the observed solar RV variability with a measured standard deviation of the residuals of $\sim0.91$ m/s. Photometric simulations capture the overall variability trends, confirming that \texttt{SOAP} can reliably model the impact of both spots and faculae on solar RVs and photometry.}


\titlerunning{Modelling solar radial velocities and photometric variability with \texttt{SOAPv4}}

\authorrunning{A. Barka et al.}

\maketitle

\section{Introduction}

From the detection of the first exoplanet orbiting a solar-like star in 1995 \citep{1995Natur.378..355M}, the exoplanet research field has seen extraordinary growth, with almost 6000 exoplanets confirmed to date\footnote{\url{https://exoplanet.eu}.}. Currently, one of the primary objectives of the community is the detection and characterisation of Earth-like planets orbiting within the habitable zone of solar-like stars.  \par 
The radial velocity (RV) technique, the first successful method in exoplanet research and still among the most productive, faces particular challenges in this context. These Earth-like planets are difficult to detect due to their small size and wide orbits around solar-like stars, which induce an RV signal of only about 10 cm/s on their host stars \citep{2014Natur.513..328M}.

Lately, the RV method has seen significant advancements, first with the sub-metre-per-second precision delivered by instruments such as the         High Accuracy Radial velocity Planet Searcher (HARPS; \citealt{2000SPIE.4008..582P,2003Msngr.114...20M}) and its northern hemisphere counterpart, HARPS-N (\citealt{2012SPIE.8446E..1VC}), and more recently with the advent of new-generation spectrographs such as the Echelle SPectrograph for Rocky Exoplanets and Stable Spectroscopic Observations (ESPRESSO; \citealt{refId0}), which pushes RV precision to an even higher level, reaching $\sim 10-20$ cm/s. Reaching the centimetre-per-second regime makes it possible to detect the RV signals produced by Earth-like planets. As a result, the main limitation to planet detection is no longer instrumental, but astrophysical. In high-resolution spectroscopy, a planet orbiting a solar-like star is not the only element influencing the stellar spectrum.  Intrinsic stellar phenomena, such as oscillations, granulation, and active regions, can produce signals in the spectra that mimic or hide the expected signature of an exoplanet \citep[e.g.][]{QrefId0,2012MNRAS.421L..54C, 2021arXiv210714291C, CRPHYS_2023__24_S2_205_0}.

These phenomena are all connected to stellar convection or its inhibition by magnetic activity, and they dominate on different timescales. Oscillations, acoustic waves excited by convection, dominate on timescales of a few minutes for solar-like main-sequence stars, and they can usually be averaged out with appropriate observational strategies \citep[e.g.][]{2011A&A...525A.140D,2014A&A...572A..48R, 2019LRSP...16....4G}. Granulation, a direct manifestation of convection at the stellar surface, dominates at longer timescales, from a few minutes to hours \citep[e.g.][]{2011A&A...525A.140D,2015A&A...583A.118M}, and induces a net convective blueshift effect on the integrated solar spectrum \citep{1981A&A....96..345D}. This effect arises from the dominance of the upward flowing motion over the downward flowing material in the convective cells \citep[e.g.][]{2014masu.book.....P, 2019ApJ...879...55C}. 
On timescales of days and longer, the variability is instead dominated by active regions, i.e. localised concentrations of magnetic fields \citep{2010A&A...512A..38L, 2010A&A...512A..39M, 2011A&A...527A..82D}. Convection, among other processes, contributes to the formation and concentration of these magnetic fields. Once formed, active regions locally inhibit convection, which has two key effects \citep{2018A&A...610A..52B}. First, the suppression of convective motions inhibits the convective blueshift \citep[e.g.][]{2010A&A...512A..39M}. Second, the modified convective energy transport alters the emergent flux, producing brightness inhomogeneities across the rotating stellar surface. These appear observationally on the photosphere of solar-like stars as dark spots and bright faculae.
Spots are cooler, darker regions of the photosphere, often occurring in groups and typically associated with opposite magnetic polarities \citep{2014masu.book.....P}. They are generally composed of a darker central umbra surrounded by a lighter penumbra, both of which are cooler than the surrounding photosphere due to the local suppression of convection \citep[e.g.][]{article}. Faculae, on the other hand, are bright magnetic regions that cover a larger fraction of the stellar disc than spots, but with less contrast in the visible band. Their contrast increases towards the limb, making them more prominent at inclined viewing angles \citep{2004soas.book.....F}.
As these regions form and evolve, they break the rotational Doppler balance between the approaching and receding stellar hemispheres, leading to additional RV shifts \citep[e.g.][]{2010A&A...512A..38L,2010A&A...512A..39M}. 
\par In recent years, numerous studies have been conducted to deepen the community's understanding of these intriguing phenomena from exoplanetary signals \citep[e.g.][]{2019ApJ...874..107M,Haywood_2022,Palumbo_2024,2024A&A...687A.303M}. However, no method currently exists that allows for the unambiguous detection of the signature of an Earth-like planet orbiting a distant star. Owing to its proximity, which enables high-spatial-resolution studies, and its well-established centre of motion, the Sun is recognised by the community as the most suitable proxy for advancing towards this goal \citep{2010A&A...512A..38L, 2025Msngr.194...21S}. 

In this work we focused on the role of active regions and simulated both their RV and photometric signatures in the solar spectrum over a timescale of three years, partially covering the declining phase of solar cycle 24. We included the photometric variability since, in addition to inducing RV shifts, the flux imbalance caused by active regions also produces measurable brightness variations. We identified the active regions in solar Solar Dynamics Observatory (SDO; \citealt{2012SoPh..275....3P}) images and simulated their effect using version 4 of \texttt{Spot Oscillation and Planet} (\texttt{SOAPv4}; \citealt{refId01, Dumusque_2014,Cristo_2025}), testing its performance. We then compared our results with RV data from the HARPS-N solar telescope and with photometric data from Variability of Solar IRradiance and Gravity Oscillations (VIRGO)/Sun PhotoMeters (SPM; \citealt{1995SoPh..162..101F,1997SoPh..170....1F,2002SoPh..209..247J}).
The paper is organised as follows: Section~\ref{data} describes the SDO/Helioseismic and Magnetic Imager (HMI) and Atmospheric Imaging Assembly (AIA) datasets used for the identification of active regions, together with the RV and photometric data employed for comparison. Section~\ref{active_region} details the identification process, while Sect.~\ref{soap} presents the simulations performed with \texttt{SOAPv4}. Section~\ref{results} reports and discusses our final results, comparing the simulation outputs obtained with the different identification methods. Finally, Sect.~\ref{summary} summarises our conclusions.

\section{Data}
\label{data}
\subsection{SDO images}
We identified solar active regions using full-disc observations from SDO \citep{2012SoPh..275....3P}. Sunspots were detected using continuum `intensitygrams' from HMI \citep{2012SoPh..275..229S}, which observes the solar photosphere near the FeI 6173 \AA\ absorption line. For facular identification, we used both HMI magnetograms, which trace photospheric magnetic fields, and ultraviolet (UV) images from AIA \citep{2012SoPh..275...17L}, specifically the 1700 \AA\ continuum. This channel probes the upper photosphere and lower chromosphere \citep{Simoes_2019}, where faculae extend into plages. At this wavelength, the intensity contrast of bright regions increases, which facilitates their detection, while the resulting area measurements remain in reasonable agreement with the underlying photospheric faculae \citep{shapir}. In the left panels of Fig.~\ref{fig:fitsfiles}, we show an example of the HMI and AIA images used for our work.
We downloaded the images from the MEDOC/IAS\footnote{\url{https://idoc-medoc.ias.u-psud.fr/sitools/client-user/index.html?project=Medoc-Solar-Portal}.} (Multi Experiment Data \& Operation Center at Institut d'Astrophysique Spatiale in Orsay) archive, using the PySitool2 client. For the HMI data, we selected the 720s cadence series, and for the AIA images, we used the level 1 data with a 24s cadence.
\subsection{The Debrecen catalogue}
In addition to our own identification, as spot data input for the \texttt{SOAP} simulations, we also used the Debrecen Heliographic Catalogue \citep{2016SoPh..291.3081B, 10.1093/mnras/stw2667}. This catalogue provides daily sunspot positions and areas derived from white-light full-disc observations obtained at the Heliophysical Observatory of the Hungarian Academy of Sciences (Debrecen, Hungary), its Gyula Observing Station, and several other observatories. When ground-based observations were unavailable, quasi-continuum images from the Michelson Doppler Imager (MDI) on board the Solar and Heliospheric Observatory (SOHO) were used instead.

\subsection{HARPS-N RVs}
The HARPS-N solar telescope \citep{Dumusque_2015, 2016SPIE.9912E..6ZP}, mounted on the 3.58 m Telescopio Nazionale \textit{Galileo} (TNG) at the Roque de los Muchachos Observatory, La Palma, has been acquiring Sun-as-a-star observations since 2015 with a 5-minute cadence and sub-metre-per-second precision. The available dataset covers the period from 14 July 2015 to 15 July 2018 \citep{2019MNRAS.487.1082C}, overlapping with the declining phase of solar cycle 24.

We used RVs that were obtained from the cross-correlation functions (CCFs) between the solar spectra and spectral line masks with associated line weights, following standard procedures used in high-resolution spectrograph pipelines \citep{1996A&AS..119..373B,2002A&A...388..632P}. The CCFs were computed from three different independent pipelines: (i) the official HARPS-N Data Reduction Software (DRS)\footnote{\url{https://dace.unige.ch/}.} version 3.0.1, (ii) \texttt{YARARA} \citep{2021A&A...653A..43C,2023A&A...678A...2C}, and (iii) \texttt{ARVE}\footnote{\url{https://github.com/almoulla/arve}.} (Analyzing Radial Velocity Elements, \citealt{2025A&A...701A.266A}). There are a few key differences between these different RV extractions. The official DRS computes the CCFs on the observed spectra (which are not corrected for telluric absorption) using a pre-computed spectral type-dependent line mask that excludes wavelength regions known to have strong telluric features. \texttt{YARARA} performs sequential spectral-level corrections for telluric absorption lines, instrumental systematics and stellar activity before extracting CCF or line-by-line (LBL) RVs; in \texttt{YARARA}, the line mask is built by identifying local minima in a master stellar spectrum built from all observations. \texttt{ARVE} is an open-source Python package that supports multiple RV extraction techniques, including CCF and LBL. In \texttt{ARVE}, the line mask is built from queried Vienna Atomic Line Database (VALD; \citealt{2015PhyS...90e4005R}) line lists for various spectral types. Unlike \texttt{YARARA}, \texttt{ARVE} does not perform spectral modifications; however, it can accept various spectral formats, including those outputted from \texttt{YARARA}. The \texttt{ARVE} line mask can be adapted to include or exclude telluric wavelength regions depending on whether that correction has been previously applied to the spectra. 
For the DRS RVs, we only selected RV data for which the airmass was below $2.25$ and the quality flag\footnote{Interpreted as the probability of an observation not belonging to outliers caused by poor weather conditions, as described in \cite{2019MNRAS.487.1082C}.} was above $0.95$, computing the daily weighted averages. Regarding \texttt{YARARA} and \texttt{ARVE}, we focused on the RVs derived from spectra corrected for tellurics, instrumental systematics and stellar activity, with stellar activity re-injected. \texttt{YARARA} corrects for stellar activity by performing a multi-linear regression on the flux time series at each sampled wavelength, using common stellar activity proxies such as the CCF contrast and the full width at half maximum. While \texttt{YARARA} can deliver spectra corrected for stellar activity, we also made use of datasets in which the stellar activity correction has been intentionally reinjected, which allowed us to study the impact of stellar activity while correcting all the other factors. For both pipelines, we thus constructed a daily cadence RV time series by adopting the mean value per day.

For every valid HARPS-N day, we retrieved a single HMI continuum, HMI magnetogram, and AIA 1700 \AA\ image closest to the median HARPS-N timestamp. If any of these SDO products were unavailable within $\pm 2$ hours of that time, the day was excluded.
This procedure yielded daily datasets for each SDO instrument, resulting in about 534 days of matched observations.

\subsection{VIRGO/SPM photometric data}
For the photometric dataset, we used observations from the VIRGO/SPM instrument \citep{1995SoPh..162..101F, 1997SoPh..170....1F,2002SoPh..209..247J} on board SOHO. The SPM consists of three independent photometric channels, each with a bandwidth of 5 nm, centred at 402 nm (blue), 500 nm (green), and 862 nm (red), and provides measurements with a cadence of one minute. In this work, we focused on the green channel, as it lies near the centre of the HARPS-N spectral range (380–690 nm) and achieves sub-parts-per-million photometric precision.

\section{Active-region identification}
\label{active_region}
We performed the identification of active regions on the Sun's surface in SDO images through a code based on mathematical morphological (MM) transforms developed by \citet{BARATA201870} and \citet{2020A&C....3200385C}. The software identifies spots and plages in each observation and derives key information, namely the number of spots/plages, their latitude and longitude in Heliographic coordinates, and their area, providing both projected and corrected for foreshortening measurements. Mathematical morphology is a theory for image processing, based on set theory, which consists of comparing geometrical structures in an image with a predefined shape, known as the structuring element, to extract information about specific patterns. Originally developed for binary images and later extended to grey-scale and 3D images, it relies on a set of operators that can be applied individually or sequentially to produce increasingly sophisticated transformations. Its flexibility and ability to handle complex geometries make MM particularly well-suited for the identification of solar spots and faculae \citep{Bourgeois2024SunspotsIT}.
In this section we describe the steps and operators involved in the active-region identification procedure.

\begin{figure*}[]
\centering
\begin{subfigure}[t]{0.45\textwidth}
    \centering
    \includegraphics[width=0.85\textwidth, trim=0.4cm 0.4cm 0.4cm 0.4cm, clip]{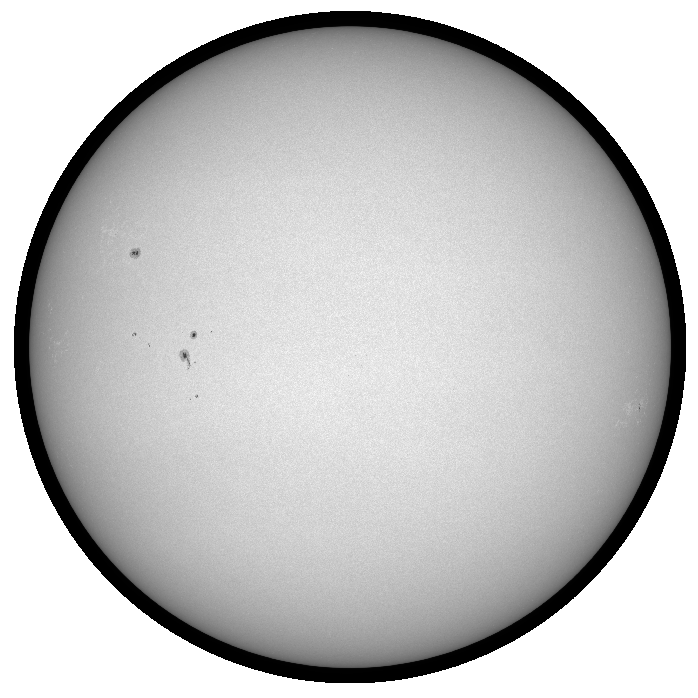}
    \caption{}
\label{a1}
\end{subfigure}
\hfill
\begin{subfigure}[t]{0.45\textwidth}
    \centering
        \rotatebox[origin=c]{180}{\reflectbox{\includegraphics[width=0.85\textwidth]{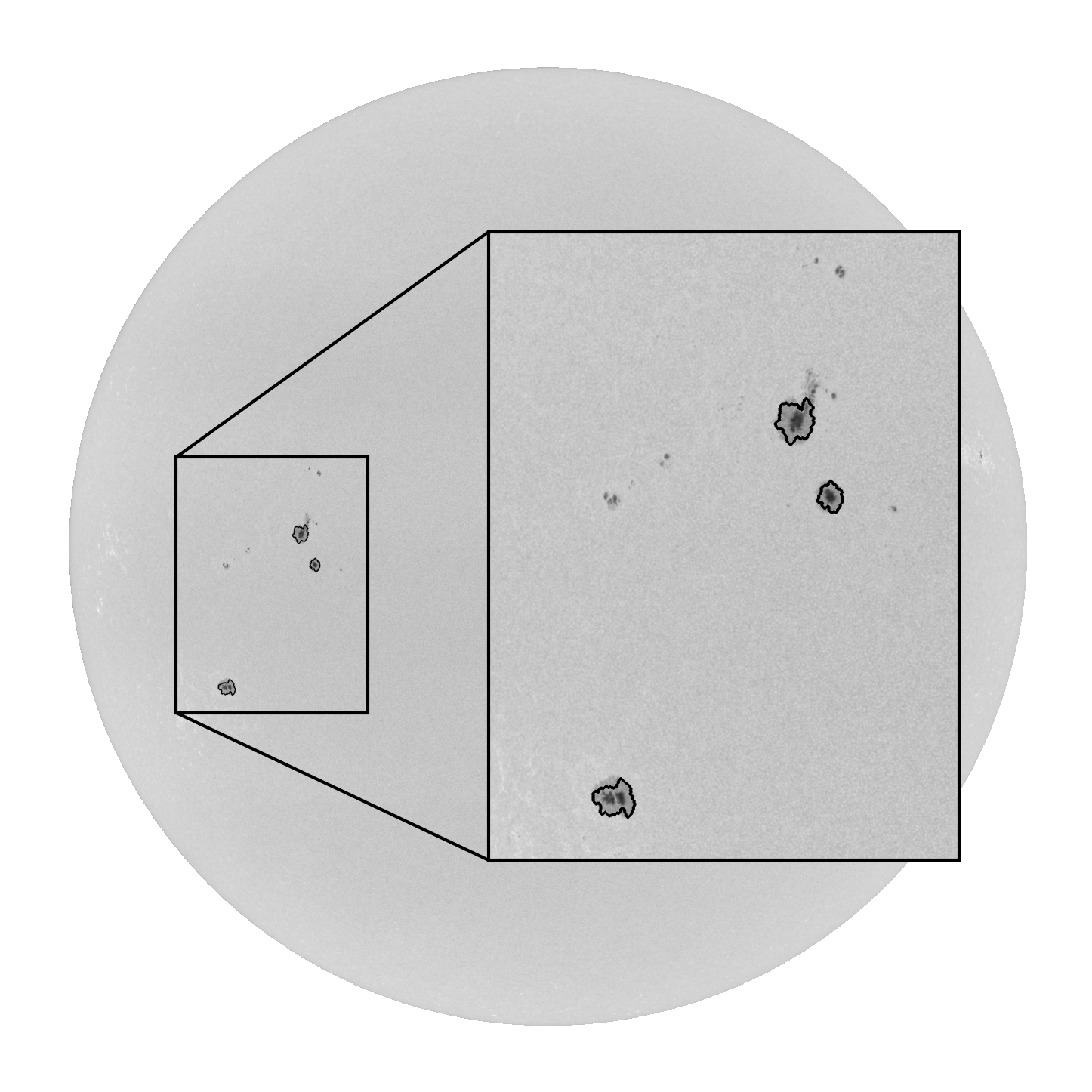}}}
    \caption{}
    \label{b1}
\end{subfigure}
\begin{subfigure}[t]{0.45\textwidth}
    \centering
    \includegraphics[width=0.85\textwidth, trim=8.1cm 5cm 7.2cm 5.2cm, clip]{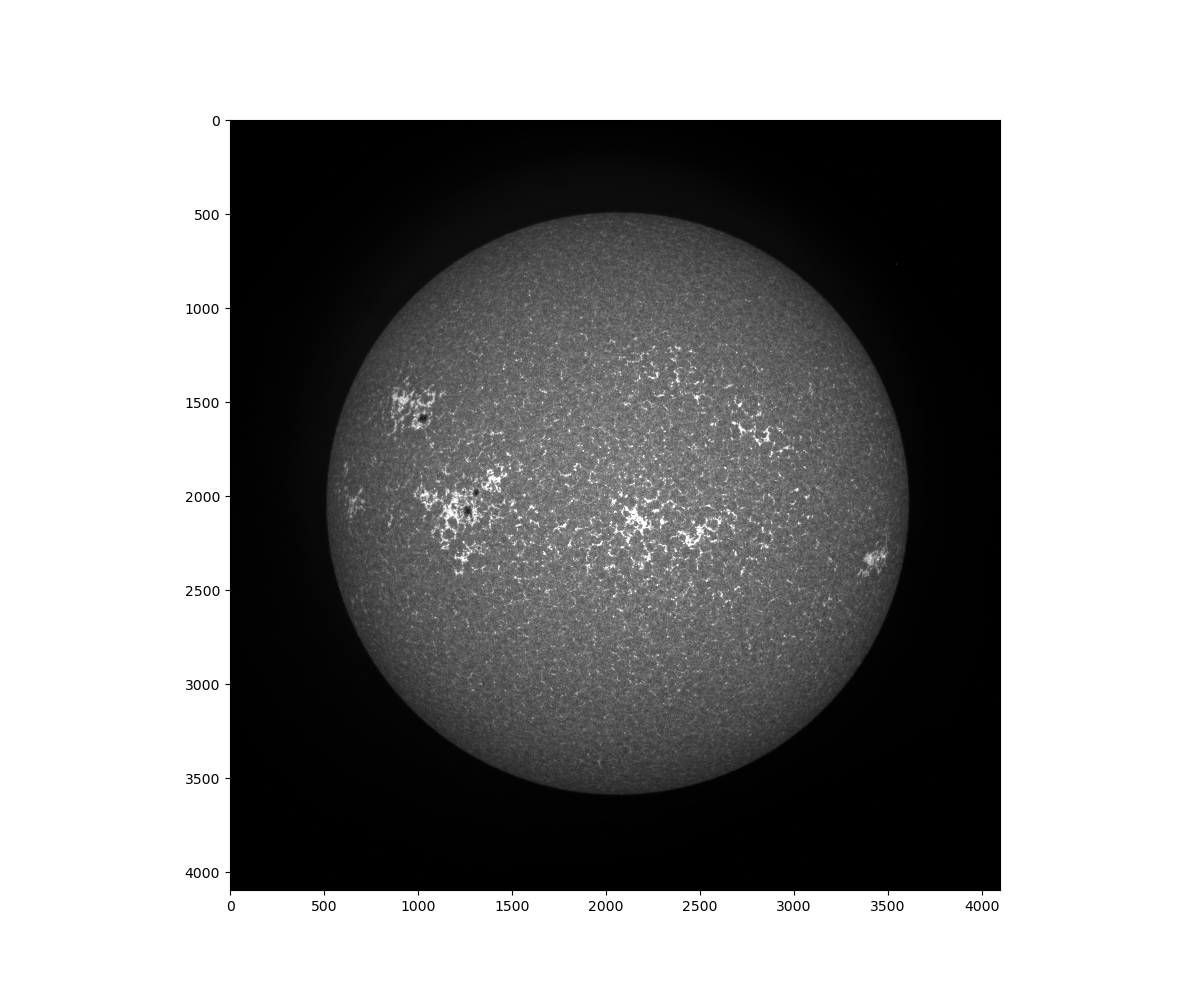}
    \caption{}
    \label{c1}
\end{subfigure}
\hfill
\begin{subfigure}[t]{0.45\textwidth}
    \centering
    \includegraphics[width=0.85\textwidth, trim=8.1cm 5cm 7.2cm 5.2cm, clip]{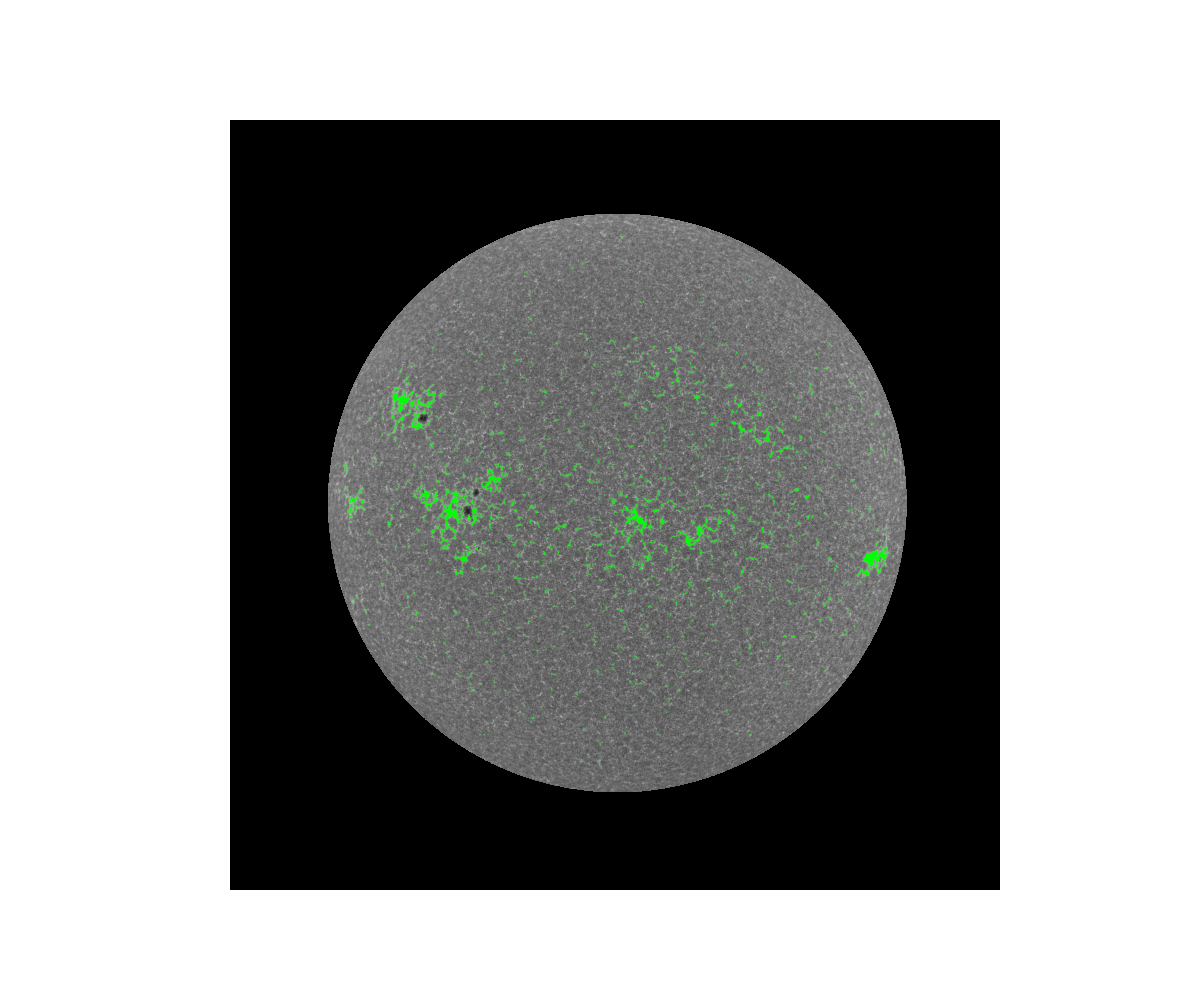}
    \caption{}
    \label{d1}
\end{subfigure}
\begin{subfigure}[t]{0.45\textwidth}
    \centering
    \includegraphics[width=0.85\textwidth, trim=6.3cm 3.2cm 5.3cm 3.4cm, clip]{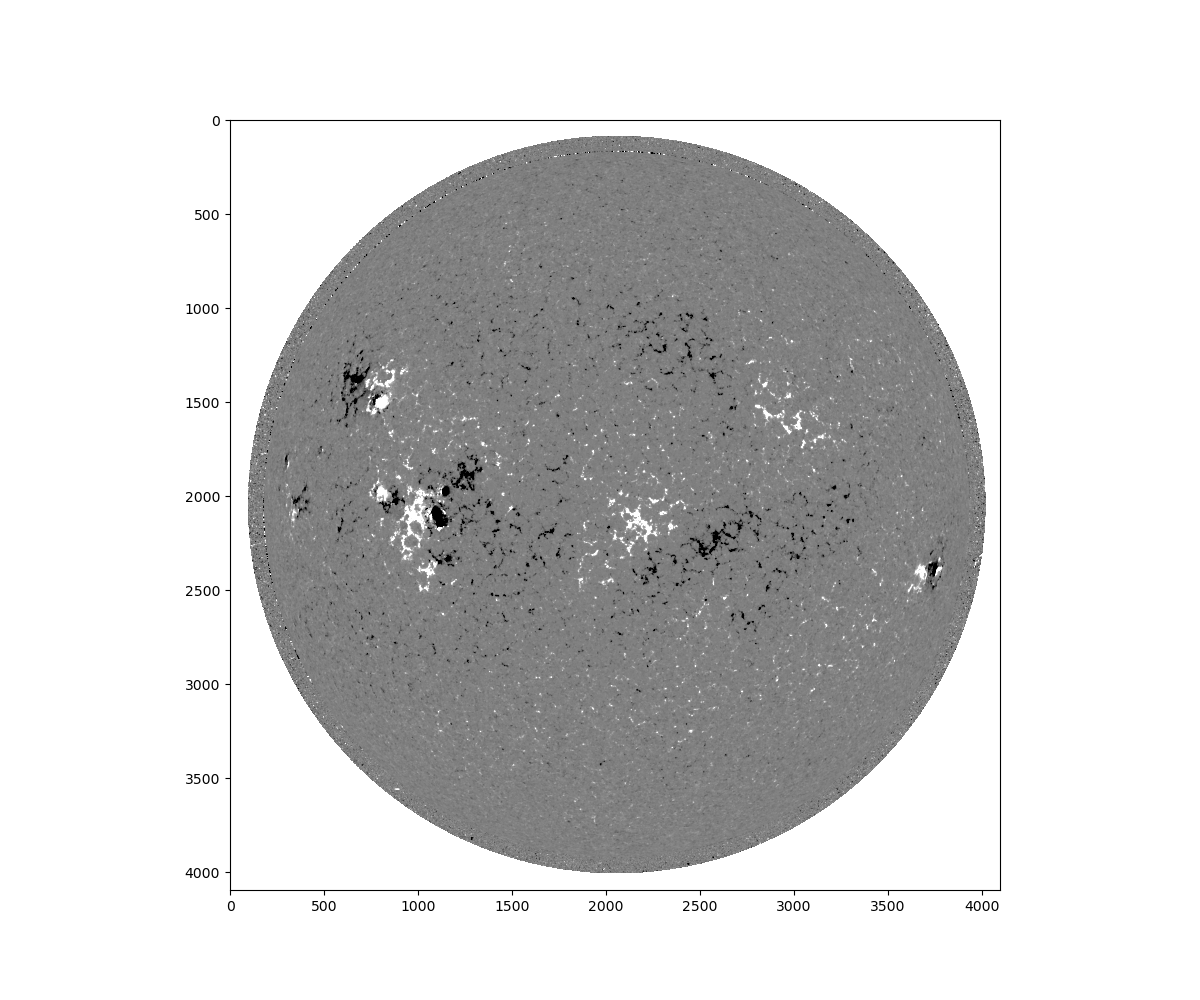}
    \caption{}
    \label{e1}
\end{subfigure}
\hfill
\begin{subfigure}[t]{0.45\textwidth}
    \centering
\rotatebox[origin=c]{180}{\reflectbox{\includegraphics[width=0.8\textwidth,trim=6.4cm 3.4cm 5.55cm 3.6cm, clip]{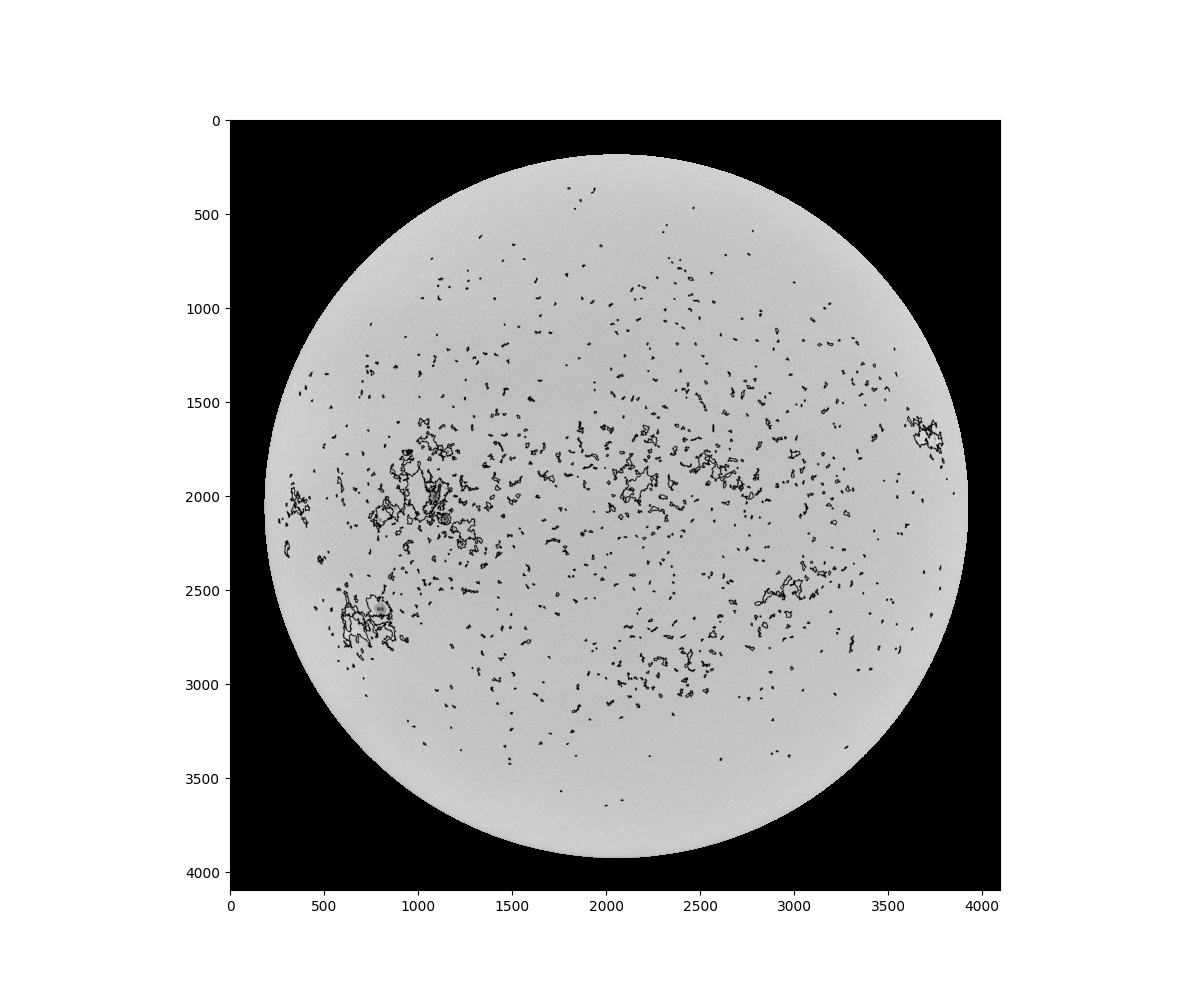}}}
    \caption{}
    \label{f1}
\end{subfigure}
\caption{Left column: Example of original SDO images from 11 August 2016: (a) HMI intensitygram, (c) HMI magnetogram, and (e) AIA UV image. 
Right column: Corresponding active-region identifications for the same observation: (b) HMI intensitygram, (d) HMI magnetogram, and (f) AIA UV image. The contours mark the detected active regions.}
\label{fig:fitsfiles}
\end{figure*}

\subsection{Correction and normalisation of images}
\label{img_corr}
Before proceeding with the identification process, we corrected the HMI intensitygrams and the AIA images for the limb darkening (LD). As a first step, we redefined the pixel coordinates relative to the solar disc centre, allowing us to compute the angular distance of each pixel. To determine the disc centre, a binary mask of the solar disc was created for each image using an intensity threshold. A threshold of 0 was adopted for HMI images and 200 for AIA images, values chosen to reliably select all pixels within the solar disc while excluding the background. We then processed the disc contours using the \texttt{cv2.findContours} function from \texttt{OpenCV}\footnote{\url{https://github.com/opencv/opencv}.}, and computed the centre of the solar disc from the resulting contours using image moments with \texttt{cv2.moments}. The solar radius was taken from the image header, so that pixels lying outside the visible disc were automatically excluded when converting pixel positions into solar coordinates. With the disc pixels identified, we proceeded to correct the intensities by the LD using a quadratic law:
\begin{equation}
    I_\text{LD} = 1 - u_1  (1 - \cos{\theta})-u_2(1 - \cos{\theta})^2,
\end{equation}
where $u_1$ and $u_2$ are the LD parameters and $\theta$ is the angular distance to the solar disc centre for each pixel. The LD parameters were obtained using LDTk (Limb Darkening Toolkit; \citealt{Parviainen2015}). To compute them, we assumed a uniform response function, represented by a boxcar filter over instrument-specific wavelength intervals. Because the LD differs between the visible and UV bands, we derived parameters separately for HMI and for AIA: for HMI, we selected a wavelength range of 6173 \AA\ $\pm 250$ \AA, whereas for AIA we adopted a band centred at 1700 \AA, extending from 1500 \AA\ to 1900 \AA.
The quadratic LD parameters were computed using a Monte Carlo approach with 20,000 samples, a burn-in of 500 iterations, and thinning of 25, resulting in $u_1=0.502$ and $u_2=0.162$ for HMI, and in $u_1=1.024$ and $u_2=-0.343$ for AIA. Each image was corrected by dividing it by the corresponding LD mask and normalised to its own median flux.
The resulting corrected and normalised figures are then used for the identification of dark spots (Sect~\ref{spot_id}) and bright plage (Sect.~\ref{facu_id}).

\subsection{Spot identification from HMI}
\label{spot_id}
We performed the identification of spots on HMI intensitygrams using the MM algorithm. This algorithm relies on the selection of a set of parameters, known as structuring elements (\texttt{se}), defined by their shape, size, and anisotropy, which are used in the different MM transforms. The parameters were chosen through qualitative validation: for each structuring element, we tested a range of values and evaluated the resulting masks by visual inspection. The final selection, summarised in Table~\ref{tab:params_spot} and detailed below, corresponds to the configuration that most reliably reproduces the global spot structure across a representative set of SDO images.

\texttt{Closing}. The closing operation consists of applying a \texttt{dilation} followed by an \texttt{erosion} with a disc-shaped structuring element of diameter \texttt{se2}. \texttt{Dilation} expands the boundaries of bright regions in the image, whereas \texttt{erosion} contracts them. For the HMI intensitygrams, we set \texttt{se2} to 200 pixels in size, which maximises the inclusion of the full penumbra while keeping the computational time low.

\texttt{Black top-hat}. The black top-hat transformation is obtained by subtracting the original image from the result of the closing operation. This process highlights dark features, such as sunspots, against a brighter background, making them more prominent for subsequent analysis.

\texttt{Adaptive threshold}. After applying the black top-hat transformation, we used an adaptive threshold, \texttt{t2}, to generate a binary image that isolates the spots. This step sets to 1 all pixels with intensity values higher than \texttt{t2}, and to 0 those below it. We set \texttt{t2} to 0.3, a value that optimises the detection of both umbral and penumbral regions of sunspots.

\texttt{Opening by reconstruction}. This process starts by eroding the image to remove the presence of noise, followed by iterative dilation to restore the shape of the remaining features. Setting \texttt{se3} to a disk of size 11 pixels provided a stable balance between noise removal during erosion and the preservation of real features during dilation across the tested images.
This step results in a binary image with improved identification of sunspots for the subsequent analysis.

In the final image, we extracted the contours of the identified spots with \texttt{cv2.findContours} and determined their centroids using \texttt{cv2.moments}. We computed the area of each contour using \texttt{cv2.contourArea}, which counts the number of pixels inside each contour. The measured area was then corrected for foreshortening using the following formula from \cite{2002JBAA..112..353M}: 
\begin{equation}
    A_{\text{corr}} = A_{\text{pix}} / (2\pi R_{\odot} ^2\cos{\theta}),
\end{equation}
where $A_{\text{pix}}$ is the identified area in pixels, $R_\odot$ is the radius of the solar disc in the image, and $\theta$ is the angular distance between the centroid of the identified contours and the solar disc centre.
Figures~\ref{a1} and~\ref{b1} present an example of an HMI intensitygram from 11 August 2016, where three sunspots can be seen in black, along with the corresponding processed image in which these spots are identified. Bright faculae are also visible near the solar limb.
As an alternative to the MM spot identification method, we also employed the Debrecen sunspot catalogue to obtain sunspot areas and positions, which we used to run additional validation tests.
\par

\begin{table}[h!]
  \centering
  \renewcommand{\arraystretch}{1.3}
  
  \caption{MM parameters adopted for the identification of spots (a) and plages (b); see the text for further information.}
  
  \begin{subtable}{0.8\linewidth}
    \centering
    \caption{Spot identification.}
    \begin{tabular}{l | c}
      \hline\hline
      Operation & Parameter value \\
      \hline
      Closing & \texttt{se2} = 200 \\
      Adaptive threshold & \texttt{t2} = 0.3 \\
      Opening by reconstruction & \texttt{se3} = 11 \\
      \hline
    \end{tabular}
    \label{tab:params_spot}

  \end{subtable}

  \vspace{0.5cm} 

  \begin{subtable}{0.8\linewidth}
    \centering
    \caption{Plage identification.}
    
    \begin{tabular}{l | c}
      \hline\hline
      Operation & Parameter value \\
      \hline
      Opening & \texttt{se2} = 200 \\
      Adaptive threshold & \texttt{t2} = 1 \\
      Opening by reconstruction & \texttt{se3} = 3 \\
      \hline
    \end{tabular}
    \label{tab:params_plage}

  \end{subtable}

\end{table}

\subsection{Plage identification from AIA 1700 \AA}
\label{facu_id}
To identify plages and determine their properties, we used AIA UV images. As described below, the steps to reach the plage identification and the order they are applied do not differ from those of the spot identification. The parameters chosen for the plages case are summarised in Table~\ref{tab:params_plage}.

\texttt{Opening}. As a first step, we applied an opening operation, which is the reverse of the closing used for spots, to highlight bright features. This involves an erosion followed by a dilation, setting \texttt{se2} to 200.

\texttt{White top-hat}. By subtracting the result of the opening operation from the original image, we performed a white top-hat transformation to enhance bright features against the darker background.

\texttt{Adaptive threshold}. A binary image is then created by applying a threshold with \texttt{t2} equal to 1, setting all pixels with intensity above this value to 1, while the rest are set to 0, isolating the brightest regions.

\texttt{Opening by reconstruction}. Finally, we applied an opening by reconstruction, setting \texttt{se3} to 3, removing small-scale noise while preserving the structure and contours of the identified plages.
Figures~\ref{c1} and~\ref{d1} present an example of an AIA image from 11 August 2016, where bright plages are visible, along with the corresponding processed image in which these plages are identified in green. The three sunspots previously detected are also visible.

\subsection{Facula identification from HMI magnetograms}

\begin{figure*}[]
\centering

    \begin{subfigure}[t]{0.5\textwidth}
        \centering
        {\includegraphics[width=0.9\textwidth, trim=0.6cm 0.6cm 0.6cm 0.6cm, clip]{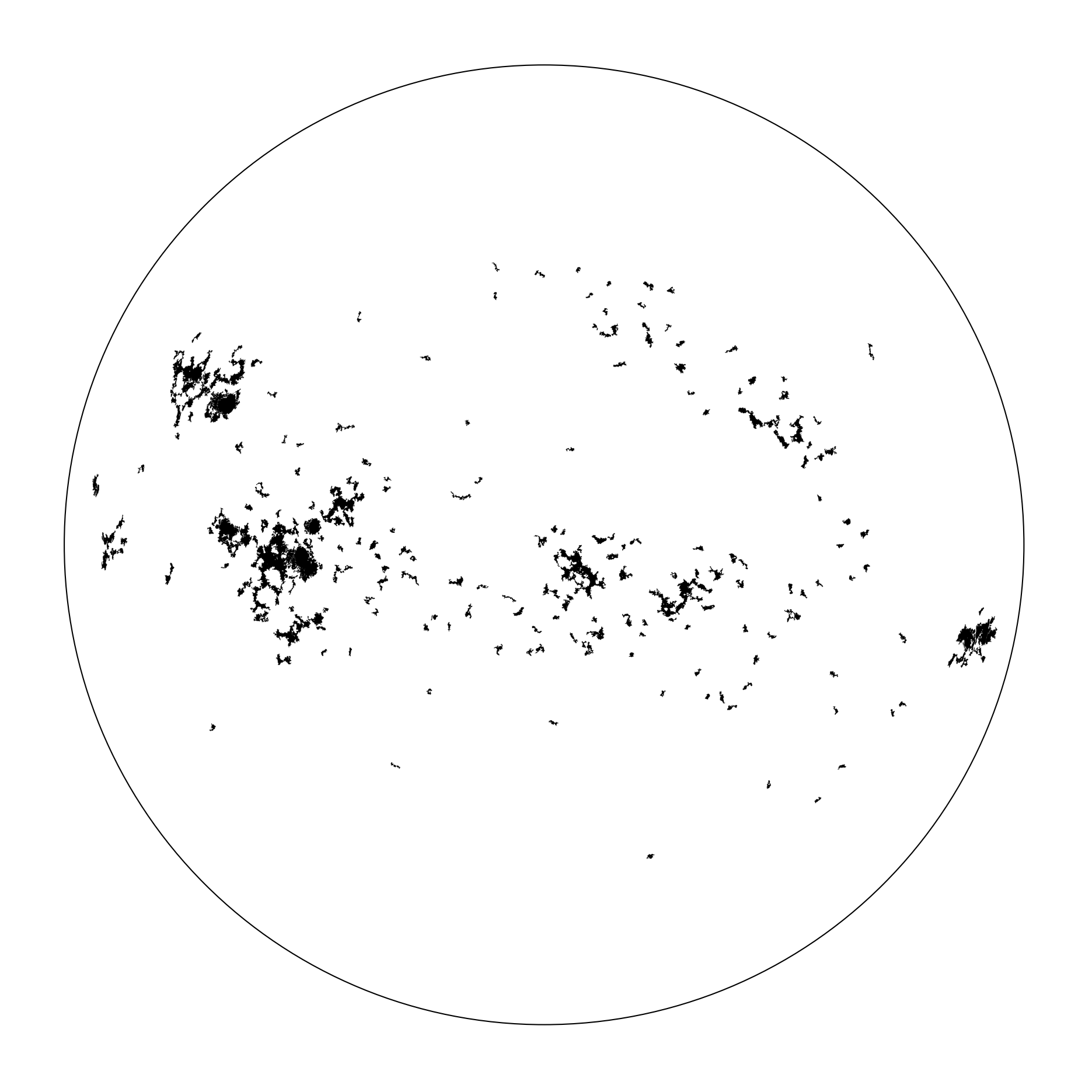}}
        \caption{}
    \label{a}
    \end{subfigure}%
    \hfill
    \begin{subfigure}[t]{0.5\textwidth}
        \centering
            \rotatebox[origin=c]{180}{\reflectbox{\includegraphics[width=0.9\textwidth, trim= 2.2cm 2.2cm 2.2cm 2.2cm, clip]{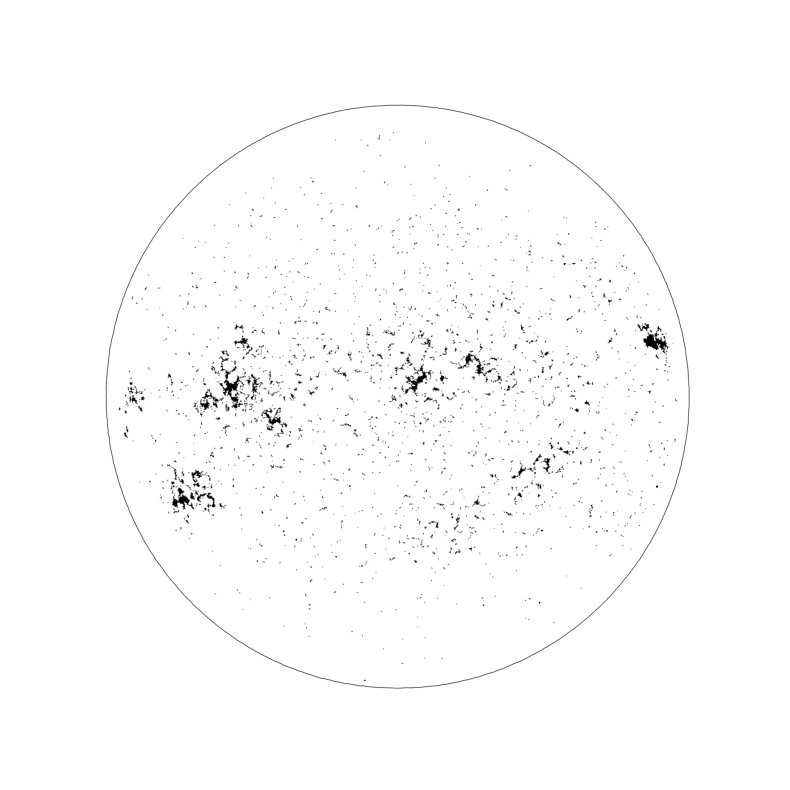}}}
        \caption{}
        \label{b}

\end{subfigure}

\caption{Left: Larger facular regions identified using the H16 method. Right: Plages detected using the MM method. The example is from 11 August 2016.}
\label{fig:faculae_H16/MM}
\end{figure*}

For comparison with our MM-based identification, we also used a second dataset based on the method described in \citet[hereafter H16]{Haywood_2016}, \citet{2019ApJ...874..107M}, and \citet{Palumbo_2024}. The H16 procedure detects both faculae and spots from HMI magnetograms and intensitygrams; however, in our work, we only used the facula identification.
In this approach, faculae are identified from HMI magnetograms and intensitygrams. Active pixels in the magnetograms are first defined as those satisfying the condition

\begin{equation}
|B_{\text{rad},ij}| > 24 ~\text{G}/\cos{\theta_{ij}} ,
\end{equation}
where $B_{\text{rad},ij}$ is the radial component of the magnetic field in pixel \texttt{ij}, due to foreshortening, and $\theta_{ij}$ is the angular distance of that pixel. The quiet-Sun reference is then determined from the LD corrected and normalised HMI intensitygrams (see Sect~\ref{img_corr}) by computing the mean intensity of the non-active pixels, denoted as $\hat{I}_\text{quiet}$. Facular pixels are subsequently classified as those that are both magnetically active and brighter than the quiet Sun, i.e. with intensity $I > 0.89~\hat{I}_\text{quiet}$ \citep{yeo}. In this dataset, only the largest facular regions were considered, referred to as plages in H16.
A comparative visualisation of the plage/faculae regions identified with the MM and H16 methods is shown in Fig.~\ref{fig:faculae_H16/MM}. While the MM method captures a larger number of small-scale features, the largest contiguous plage structures are more extended in the H16 identification, resulting in a larger total plage area and, consequently, a higher filling factor for H16.

In addition to this dataset, we constructed a second one by applying the same H16 identification criteria, followed by an opening by reconstruction setting \texttt{se3} to a disc of pixel size 6 (hereafter H16 + MM). This additional morphological step improved the definition of the facular regions and suppressed small-scale noise. Figures~\ref{e1} and~\ref{f1} show an example of an HMI magnetogram from 11 August 2016, displaying the magnetically active regions in black and white, together with the corresponding facular regions outlined by the contours.
From this processed image, the area and centroids of the identified faculae were computed using \texttt{openCV}.

\section{\texttt{SOAPv4} simulations}
\label{soap}

\begin{table}[h]
\caption{Solar input parameters for the \texttt{SOAP} simulations.}

  \centering\renewcommand{\arraystretch}{1.5}
    \begin{tabular}{c|c}\hline\hline
    Solar parameters \\ \hline
         $P_{\text{rot}}$ (days)&  $25.4 $ \\
         $i$ (degree) &  $90$ \\
         $R$ ({R$_{\odot}$}) &  $1$ \\
         $M$ (M$_{\odot}$) &  $1$   \\ 
         $T_{\text{eff}}$ (K) &  $5772 $\\ 
    \end{tabular}
    \tablefoot{The rotational period $P_{\text{rot}}$ and effective temperature $T_{\text{eff}}$ were taken from \url{https://web.archive.org/web/20250531123908/https://nssdc.gsfc.nasa.gov/planetary/factsheet/sunfact.html}.}
    \label{tab:sol_param}
\end{table}

\texttt{SOAP} \citep{refId01} is a numerical code capable of reproducing the impact of stellar activity on RVs and photometry. Specifically, we used the \texttt{SOAPv4} version (\citealt{Cristo_2025}), an improved version of \texttt{SOAP2.0} \citep{Dumusque_2014}. The simulations use two spectra \citep{1998assp.book.....W,2005asus.book.....W} obtained with the Fourier Transform Spectrograph at the Kitt Peak Observatory as input: one for the quiet photosphere at the disc centre and one for a spot umbra, which is used as representative for both the dark spots and bright faculae that define the active regions. The code performed the simulations using CCFs constructed from the aforementioned spectra with a G2 HARPS template.
To construct the time series of disc-integrated CCFs, the code starts by dividing the solar disc into a user-defined grid of $N$ cells with equal projected area. Each cell is assigned one of the two CCFs depending on whether it corresponds to a quiet or active region. The velocity of the CCF assigned to each cell is shifted according to the cell’s position on the disc and the solar rotational period $P_{\text{rot}}$. The shifted CCFs are then weighted by the LD of each cell and combined across the disc to produce the final integrated stellar CCF from which the RV is extracted. The photometric signal is obtained analogously, by combining the intensities of all cells across the solar disc, weighted by their LD and according to whether they are within the quiet photosphere or active regions.
We modified the code to automate the input of spot and facula properties for each observation, feeding the derived spot and facula parameters from Sect.~\ref{active_region}, and producing a simulated RV and photometric variation for each time step.

For our simulations, we set a grid size of $3000 \times 3000$, which corresponds to a resolution of 0.12 arcsec per pixel at the disc centre. Although this is finer than the native HMI resolution, the MM method does not detect very small-scale dark features, as illustrated in the top panels of Fig.~\ref{fig:fitsfiles}. When mapped onto the \texttt{SOAPv4} grid, the smallest detected spots are typically sampled by about 5 pixels, ensuring that the adopted grid density does not introduce unresolved structures or affect the results.
By default, \texttt{SOAP} assumes a circular shape for the spot or facula at the disc centre and then shifts it to the required position, defined by the spot's or facula's latitude and longitude, applying the corresponding area projection. The size of each feature is expressed as the ratio of the feature's radius to the solar disc radius. 
We also specified the temperature contrast relative to the solar surface, following the prescription of \citet{2010A&A...512A..39M}. In particular, we adopted a temperature contrast of $-663$~K for spots, while for faculae we used a variable contrast, $\Delta T_\mathrm{F}$, defined through the following parametrisation described in \citet{2010A&A...512A..39M}:
\begin{equation}
    \Delta T_{\text{F}} = 250.9 - 407.7 \cos{\theta} + 190.9 \cos^{2}{\theta}.
\end{equation}
In addition, \texttt{SOAPv4} assumes total inhibition of the convective blueshift within spots and faculae, with an amplitude of 350 m/s obtained from the spot observation at the centre of the disc \citep{Dumusque_2014}. The amplitude is subsequently scaled as a function of disc position to incorporate projection effects.

To perform the simulations, \texttt{SOAP} requires stellar parameters, as well as spot and facula properties. The solar parameters adopted in this work are listed in Table~\ref{tab:sol_param}. The LD parameters are obtained from LDTk, as described in Sect.~\ref{img_corr}, within the visible band (380-690 nm). We specifically selected this band because
it corresponds to the spectral range observed by HARPS-N and VIRGO/SPM. 

As mentioned above, we simulated the photometric time series with \texttt{SOAPv4}, in addition to RVs, and compared them with observations from the VIRGO/SPM instrument. We focused on the green channel, centred at 500 nm, as it closely corresponds to the central wavelength range of HARPS-N. To ensure consistency with the VIRGO/SPM data processing, we subtracted a rolling mean from the simulated photometry. We chose a window size of 200 data points for the rolling mean in order to smooth out long-term trends while ensuring that variability at the rotational timescale is preserved.

\section{Results and discussion}
\label{results}
In this section we present the results obtained from Sects.~\ref{active_region} and~\ref{soap}. In Sect.~\ref{ffcomp} we compare our filling factors with other catalogues. In Sects.~\ref{rvcomp} and~\ref{photocomp} we compare our simulations obtained by combining the filling factors and \texttt{SOAPv4} simulations, with Sun-as-a-star observational data from HARPS-N and VIRGO/SPM.

\subsection{Filling factor comparison}
\label{ffcomp}
\begin{figure*}[t]
\centering

   \begin{minipage}{0.95\textwidth}
     \centering

     \begin{subfigure}{0.46\textwidth}
         \includegraphics[width=\linewidth]{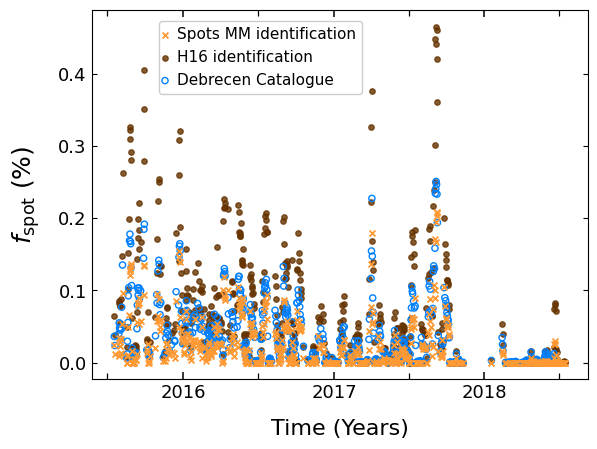}
         \caption{}
         \label{ffspottot}
     \end{subfigure}
     \hfill
     \begin{subfigure}{0.26\textwidth}
         \includegraphics[width=\linewidth]{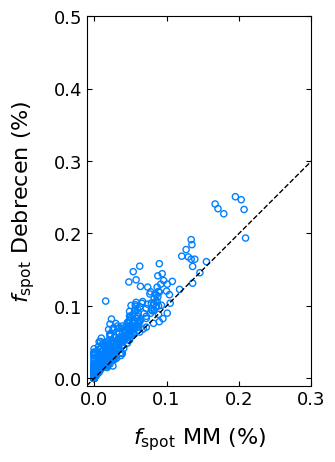}
         \caption{}
         \label{ffspotdeb}
     \end{subfigure}
     \hfill
     \begin{subfigure}{0.26\textwidth}
         \includegraphics[width=\linewidth]{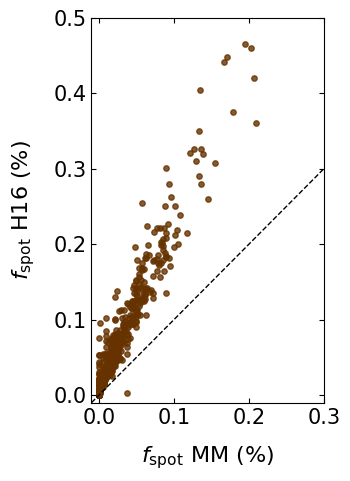}
         \caption{}
         \label{ffspotH16}
     \end{subfigure}

   \end{minipage}

\caption{Comparison of the spot filling factors measured from MM on HMI images (orange), from the Debrecen catalogue (blue), and following H16 (brown). Panel (a) shows the temporal evolution of each, while panels (b) and (c) show, respectively, the latter two as a function of the former, which we take as reference in our sample.}
\label{fig:fillfact}
\end{figure*}

\begin{figure*}[t]
\centering

   \begin{minipage}{0.95\textwidth}
     \centering

     \begin{subfigure}{0.49\textwidth}
         \includegraphics[width=\linewidth]{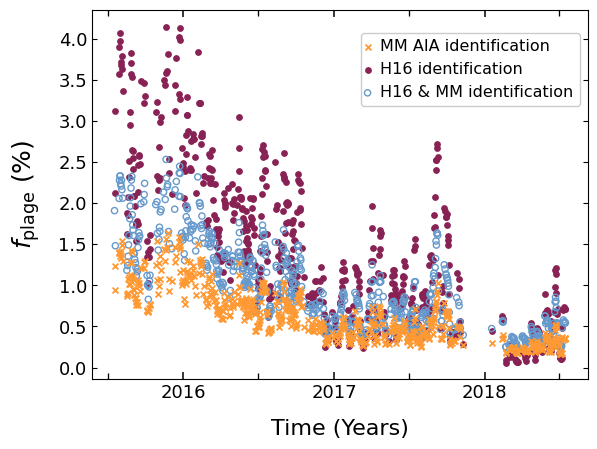}
         \caption{}
         \label{ffplagetot}
     \end{subfigure}
     \hfill
     \begin{subfigure}{0.25\textwidth}
         \includegraphics[width=\linewidth]{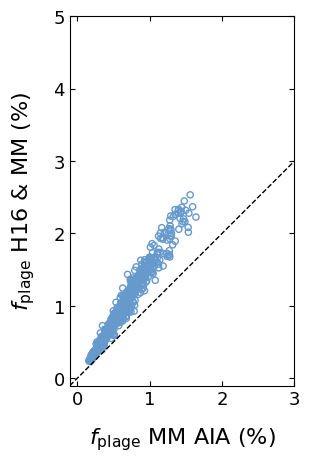}
         \caption{}
         \label{ffplageMM+H}
     \end{subfigure}
     \hfill
     \begin{subfigure}{0.25\textwidth}
         \includegraphics[width=\linewidth]{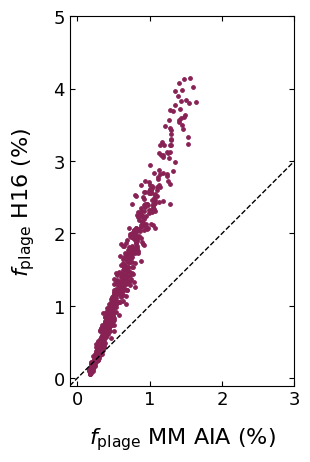}
         \caption{}
         \label{ffplageH16}
     \end{subfigure}

   \end{minipage}

\caption{Comparison of the facula filling factors measured from MM on AIA images (orange), considering H16+MM (blue), and following H16 (dark red). Panel (a) shows the temporal evolution of each, while panels (b) and (c) show, respectively, the latter two as a function of the former, which we take as reference in our sample.}
\label{fig:fillfact1}
\end{figure*}

We compared the filling factors ($f_{\text{spot}}$; $f_{\text{plage}}$) derived from the different identification methods for both spots and faculae and/or plages, shown in Figs.~\ref{fig:fillfact} and~\ref{fig:fillfact1}, respectively. Figure~\ref{ffspottot} shows the time series of $f_{\text{spot}}$. Although not used in the simulations, for completeness, we also report the $f_{\text{spot}}$ values derived using the H16 method. This approach systematically yields higher $f_{\text{spot}}$ estimates compared to both the MM identifications and the Debrecen catalogue, as shown in Fig.~\ref{ffspottot}.

The comparison between MM and Debrecen $f_{\text{spot}}$ (Fig.~\ref{ffspotdeb}) shows a tight agreement, with a slope of $\sim 1.19$. The scatter is mainly confined to lower values, where Debrecen tends to slightly overestimate the MM values. In some cases, the MM identification does not report spots listed in the Debrecen catalogue; visual inspection of these occurrences could not confirm the presence of the Debrecen spots. Nevertheless, given the small discrepancies between the MM spot dataset and the Debrecen catalogue (Fig.~\ref{ffspotdeb}), and the negligible differences observed in test simulations using both datasets, we opted to perform the RV simulations using only MM spots. This choice is further supported by the fact that spots contribute less to RV variations than faculae \citep{2010A&A...512A..39M, Dumusque_2014}. \par
In Fig.~\ref{ffspotH16}, the comparison between H16 and MM $f_{\mathrm{spot}}$ exhibits a clear linear correlation with a slope significantly larger than unity ($\sim 2.24$). All data points lie above the one-to-one relation, indicating that the H16 method systematically yields higher spot filling factors than MM, with typical values more than a factor of two larger.

To assess the sensitivity of our results to the adopted approximations, we applied our MM identification to a subset of data from 2015, consisting of 81 days. In this test, the image correction described in Sect.~\ref{img_corr} was performed using LD parameters obtained by directly fitting the HMI images, rather than computed with LDTk. In addition, the foreshortening correction was applied at the pixel level, correcting each pixel identified as part of a spot. The filling factors obtained with these implementations differ only marginally from our reference results, with variations smaller than the discrepancies found when comparing with the Debrecen catalogue. The two estimates show a strong linear correlation, with a slope of 1.03 and no significant offset, indicating an excellent level of agreement. This confirms that the approximations adopted in this work do not affect our main conclusions.

For faculae/plages, the results are presented in Fig.~\ref{fig:fillfact1}, where Fig.~\ref{ffplagetot} shows the time series of $f_{\text{plage}}$. As for the spots, the H16 method systematically overestimates the $f_{\text{plage}}$ compared to the MM identification based on AIA images. The comparison between MM AIA and the combined H16+MM identification (i.e. using the H16 method with an additional MM step, Fig.~\ref{ffplageMM+H}) shows a slope of $\sim 1.56$.
In contrast, the comparison between MM AIA and H16 $f_{\text{plage}}$ (Fig.~\ref{ffplageH16}) exhibits a steeper slope of $\sim 2.84$ and a negative offset. This offset arises because, as illustrated in Fig.\ref{fig:faculae_H16/MM}, the MM method is able to retrieve smaller-scale facular regions than H16. Hence, for time steps with small filling factors, H16 underestimates $f_{\text{plage}}$. For time steps with larger filling factors, H16 consistently yields higher $f_{\text{plage}}$ values than MM AIA.

\subsection{RV comparison}
\label{rvcomp}
For the comparison with the simulations, we took as reference the official HARPS-N DRS data product. For completeness, the results obtained with \texttt{YARARA} and \texttt{ARVE} are presented in Appendix~\ref{append}, along with a table summarising all the values for the standard deviation (std) of the residuals between simulated and observational RVs (\ref{tab:RV_std}). A normality test was also performed on the RV residuals, presented in Appendix~\ref{append2}.

Figure~\ref{fig:RVexe} presents the RV \texttt{SOAPv4} simulations based on the MM method for both spot and plage identifications, compared with the HARPS-N DRS RV measurements (left), the corresponding RV time series (top middle) together with the residual (bottom middle), and the histogram of the residuals (right).
The left panel shows that the simulated and observed RVs are generally well aligned along the 1-to-1 line, indicating that the MM-based \texttt{SOAPv4} simulations successfully reproduce the overall variability and amplitude of the measured HARPS-N RVs. 
During the gap period (around the end of 2017 and/or beginning of 2018), the fibre injecting light from the solar telescope into the calibration unit was replaced after being damaged \citep{2025A&A...693A.262Z}, which caused a systematic offset between the RVs obtained before and after the gap. To correct for this effect, we divided the dataset into pre-gap and post-gap subsets, analysed them separately, and then computed the overall std using the combined residuals for all the comparisons.
After applying this correction, the comparison between simulated and observed RVs yields a remarkably low std of $0.91$~m/s, confirming the excellent agreement between the MM-based \texttt{SOAPv4} simulations and the HARPS-N observations. This agreement is further supported by the histogram of the residuals, which is centred around zero and shows a small dispersion. 
\begin{figure*}[h]
    \centering

    \includegraphics[width=1\textwidth]{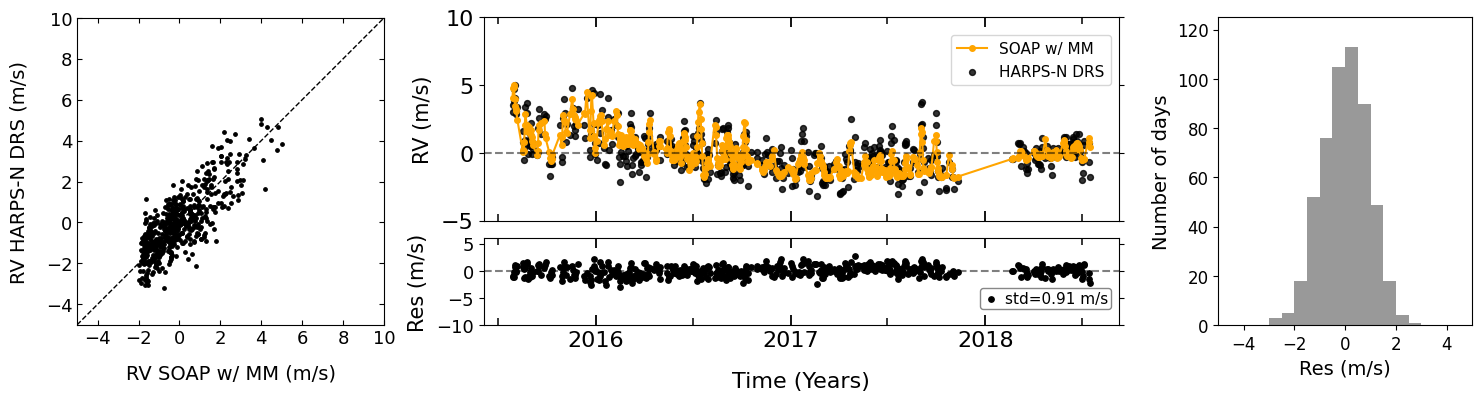}

     \caption{Left: Comparison of RVs from HARPS-N DRS and RV \texttt{SOAPv4} simulations based on the MM method for both spots and plages (\texttt{SOAP} with MM). Middle: RV time series from HARPS-N DRS (black) and from \texttt{SOAP} with MM (top), together with the residuals, calculated as the difference between observed and simulated RVs and their std value (bottom). Right: Histogram of the residuals between the two datasets.}
    \label{fig:RVexe}
\end{figure*}
The \texttt{SOAPv4} simulations based on the H16 faculae identification and the combined H16+MM approach are shown in Figs.~\ref{fig:RVH16} and \ref{RVmagneto}, respectively. In both cases, the left panels compare the HARPS-N DRS and simulated RVs. At higher RV values, the simulated points deviate increasingly from the 1-to-1 relation, indicating that both the H16 and H16+MM simulations tend to overestimate the RV amplitude during the most active phases. The corresponding time series (top panels) and residuals (bottom panels) reveal that, while these simulations reproduce more accurately the two pronounced RV peaks observed towards the end of 2017, they systematically overestimate the RV amplitude during the earlier, high-activity period, resulting in larger residuals overall. The residuals exhibit higher dispersion compared to the MM-based \texttt{SOAPv4} simulations, with stds of $1.40$~m/s for H16 and $1.54$~m/s for H16+MM. This behaviour is consistent with the filling factor analysis, where both identification methods yielded higher $f_{\text{plage}}$ values than the MM approach. Finally, the residual histograms are slightly skewed towards negative values and show broader distributions, further confirming a systematic overestimation of the observed RVs and a larger overall scatter. 
\begin{figure*}[t]
\centering

    \centering
    \includegraphics[width=\textwidth]{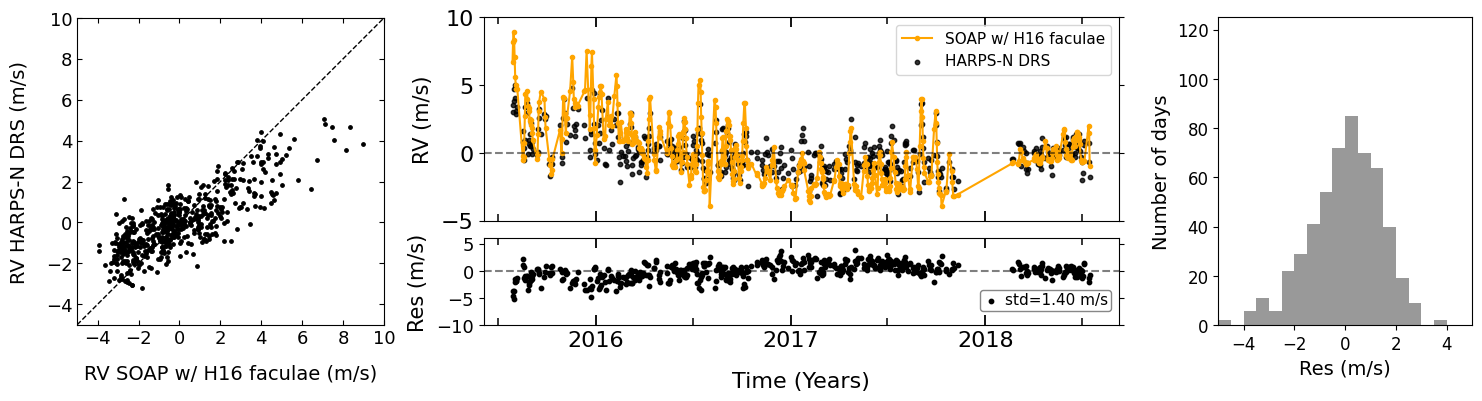}

    \caption{Same as in Fig.~\ref{fig:RVexe} but for \texttt{SOAPv4} simulations based on the H16 method.}
    \label{fig:RVH16}
\end{figure*}

\begin{figure*}[t]
\centering

    \centering
    \includegraphics[width=\textwidth]{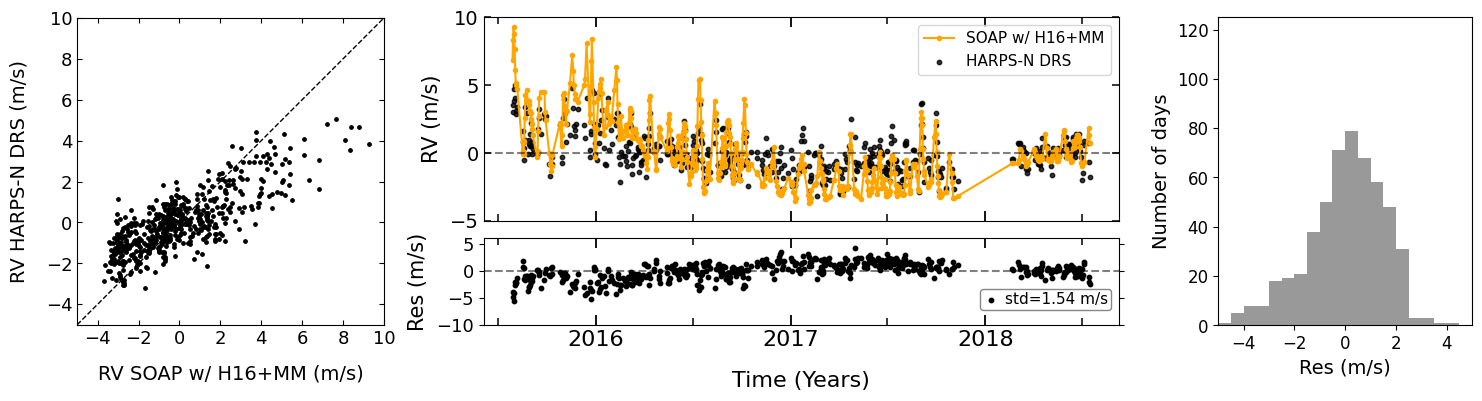}

    \caption{Same as in Fig.~\ref{fig:RVexe} but for \texttt{SOAPv4} simulations based on H16+MM.}
    \label{RVmagneto}
\end{figure*}

\subsection{Photometric variability comparison}
\label{photocomp}
\begin{figure*}[t]
    \sidecaption
    \includegraphics[width=12cm]{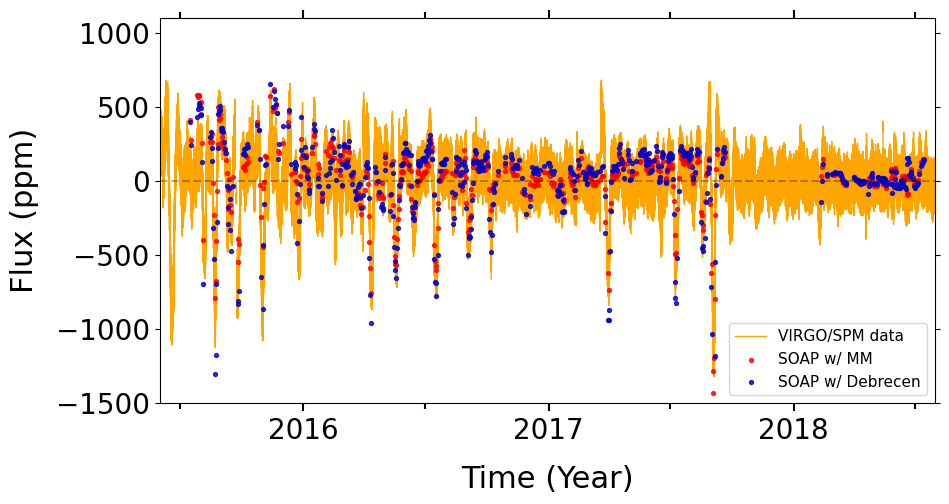}
    \caption{\texttt{SOAPv4} photometric simulation based on spots identified with MM and on spots from the Debrecen catalogue, along with faculae identified from MM, compared with photometric data from VIRGO/SPM.}
    \label{fig:virgo}
\end{figure*}

Regarding the photometric variability, the comparison of \texttt{SOAPv4} simulations and VIRGO/SPM observational data is presented in Fig.~\ref{fig:virgo}. Despite the presence of additional sources of stellar variability in the observed data, the simulations accurately reproduce both the large-scale, long-term photometric variability and the variability at the rotational timescale. A periodogram analysis of the simulated photometric time series shows clear peaks around the average solar rotation period, as well as their higher-order harmonics, confirming that the simulations capture the rotational variability. This agreement is particularly noteworthy given that the simulations incorporate only the effects of spots and faculae, while other stellar phenomena, such as granulation and oscillations, which contribute to variability in the observations, are not modelled. The strong concordance between simulations and observations is consistently observed when employing both the MM identifications and the Debrecen catalogue spots, as expected given the relatively small differences in the corresponding $f_{\text{spot}}$. However, the Debrecen-based simulation, characterised by a slightly higher $f_{\text{spot}}$, tends to overestimate the depth of the deepest photometric minima, whereas the MM-based simulation slightly underestimates them. This behaviour stems from the smaller filling factors returned by MM, which result in weaker photometric dimming. Discrepancies at the photometric maxima, on the other hand, likely reflect a combination of observational and modelling effects. The VIRGO flux is not an absolute flux measurement, being provided in ppm, so apparent peaks do not necessarily correspond to true flux enhancements relative to the quiet Sun, whose absolute level is not directly observable. In addition, facular contributions are more uncertainly modelled than spots, due to the simplified CCF representation adopted for faculae. As a result, the discrepancies are not driven by a single effect, but rather by the combination of light-curve normalisation, modelling approximations, and additional sources of brightness variability at different timescales, such as granulation and acoustic oscillations, which are not yet included in the simulations.

\section{Summary}
\label{summary}
In this work we simulated the RV and photometric signals induced by spots and faculae over a time span of three years using \texttt{SOAPv4}. To achieve this, we identified active regions in SDO images using a method based on MM transforms. To compare our identifications, we also analysed SDO magnetograms and applied the H16 method. In the case of sunspots, we further incorporated information from the Debrecen catalogue. Simulations were then compared to HARPS-N RV data and to VIRGO/SPM photometry.

Our results show that the MM identifications provide the best agreement with observations, yielding a measured RV residual std of $\sim0.91~$m/s. Faculae identified via magnetograms and the H16 method resulted in larger filling factors and higher residual stds. In the photometric domain, the simulations closely reproduce both large-scale trends and rotational timescale variability.

Despite these promising results, further refinement is needed to approach the 10 cm/s precision required to detect Earth-like planets. Within the \texttt{SOAP} framework, several limitations arise from the simplified physical description of active regions. Spots and faculae are assumed to have circular geometries, which do not fully reflect their observed morphologies, and no distinction is made between umbrae and penumbrae. In addition, the same CCF is adopted for spots and faculae, an approximation that does not capture potential differences in their spectral signatures. In this work, we focused on CCF-based simulations inherited from \texttt{SOAP2.0} primarily for computational efficiency and to assess the code’s performance at the CCF level. While \texttt{SOAPv4} enables simulations at the spectral level, which in turn enables a more detailed treatment of temperature contrasts and magnetic fields, these capabilities are not explored here. A more detailed discussion of these aspects is provided in Sect.~4 of \cite{Cristo_2025}. These limitations will be explored and progressively addressed in future work by adopting more realistic active-region morphologies, distinguishing umbrae and penumbrae, and investigating the use of full spectral information instead of CCFs. In this context, the upcoming Paranal solar ESPRESSO Telescope (PoET; \citealt{2025Msngr.194...21S}) will play a crucial role. PoET will provide disc-integrated Sun-as-a-star data and high-resolution disc-resolved data, enabling the detailed calibration of and differentiation between the signatures of the various components of active regions, such as umbrae, penumbrae, and faculae. These observations will enable a more physically motivated, realistic modelling of stellar activity in \texttt{SOAP}, particularly improving the treatment of convective blueshift inhibition and enabling spectral differentiation across active region structures.

In the future, by combining \texttt{SOAP}’s flexibility with the rich, high-quality data from PoET, we aim to significantly reduce the mismatch between models and observations. This will contribute towards a proper mitigation of stellar activity in the context of ultra-precise RV planet searches.

\begin{acknowledgements}

Funded by the European Union (ERC, FIERCE, 101052347). Views and opinions expressed are, however, those of the author(s) only and do not necessarily reflect those of the European Union or the European Research Council. Neither the European Union nor the granting authority can be held responsible for them.

This publication makes use of The Data \& Analysis Center for Exoplanets (DACE), which is a facility based at the University of Geneva (CH) dedicated to extrasolar planets data visualisation, exchange and analysis. DACE is a platform of the Swiss National Centre of Competence in Research (NCCR) PlanetS, federating the Swiss expertise in Exoplanet research. The DACE platform is available at \url{https://dace.unige.ch}.

SOHO is a project
of international collaboration between ESA and NASA. The authors thank Antonio Jim\'enez and Rafael A. Garc\'ia for providing the VIRGO/SPM data.

This work was supported by Fundação para a Ciência e a Tecnologia (FCT) through the research grant UID/04434/2025.

ARGS is supported by FCT through the work contract No.\ 2020.02480.CEECIND/CP1631/CT0001 and through the GOLF and PLATO Centre National D\'{E}tudes Spatiales grants.

KA acknowledges support from the Swiss National Science Foundation (SNSF) under the Postdoc Mobility grant P500PT\_230225.
\end{acknowledgements}

\bibliographystyle{aa} 
\bibliography{biblio} 
\onecolumn

\begin{appendix}
\section{RV residuals' normality}
\label{append2}
We evaluated whether the RV residuals are consistent with a Gaussian distribution by comparing their empirical cumulative distribution functions (CDFs) with the CDF of a normal distribution using the Kolmogorov--Smirnov (KS) test \citep{Kolmogorov33, Smirnov39}. The KS $D$ statistic measures the maximum absolute difference between the empirical and reference CDFs, with smaller $D$ statistic values indicating a closer agreement with a Gaussian distribution. The associated $p$-value quantifies the probability of obtaining a $D$ statistic value at least as large as the observed one under the null hypothesis that the residuals are normally distributed. We adopted a significance threshold of $p = 0.05$: $p$-values above this threshold indicate that the residuals are statistically consistent with a Gaussian distribution, whereas $p$-values below it indicate a significant deviation from normality.

\begin{figure*}[h]
\centering

   \begin{minipage}{0.98\textwidth}
     \centering

     \begin{subfigure}{0.34\textwidth}
         \includegraphics[width=\linewidth]{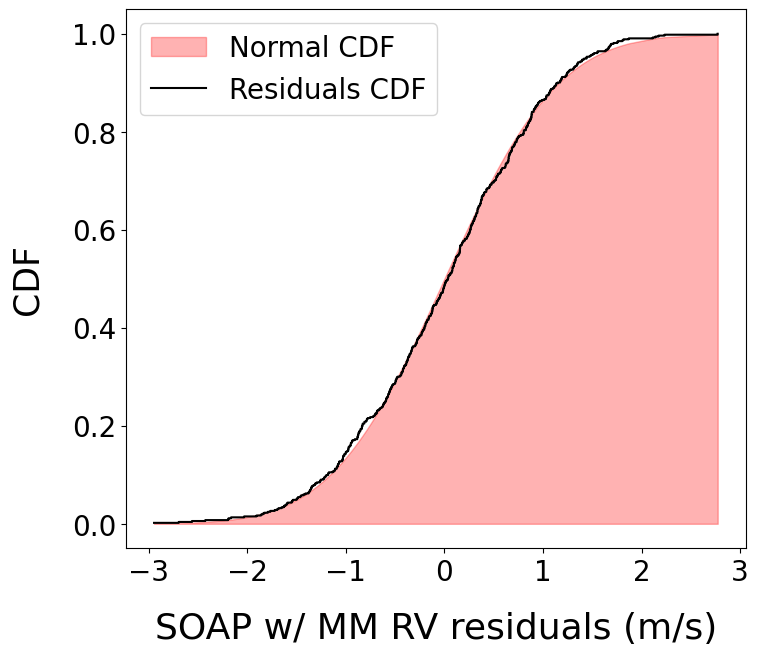}
     \end{subfigure}
     \hfill
     \begin{subfigure}{0.32\textwidth}
         \includegraphics[width=\linewidth]{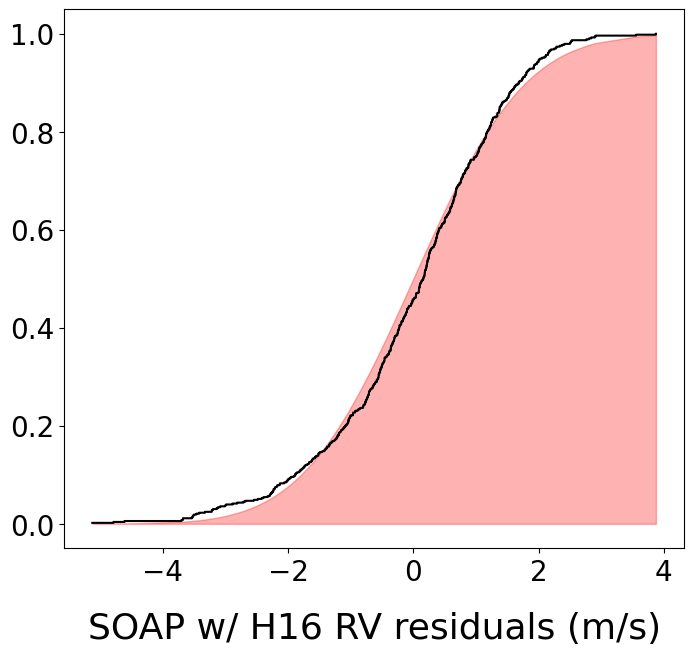}
     \end{subfigure}
     \hfill
     \begin{subfigure}{0.32\textwidth}
         \includegraphics[width=\linewidth]{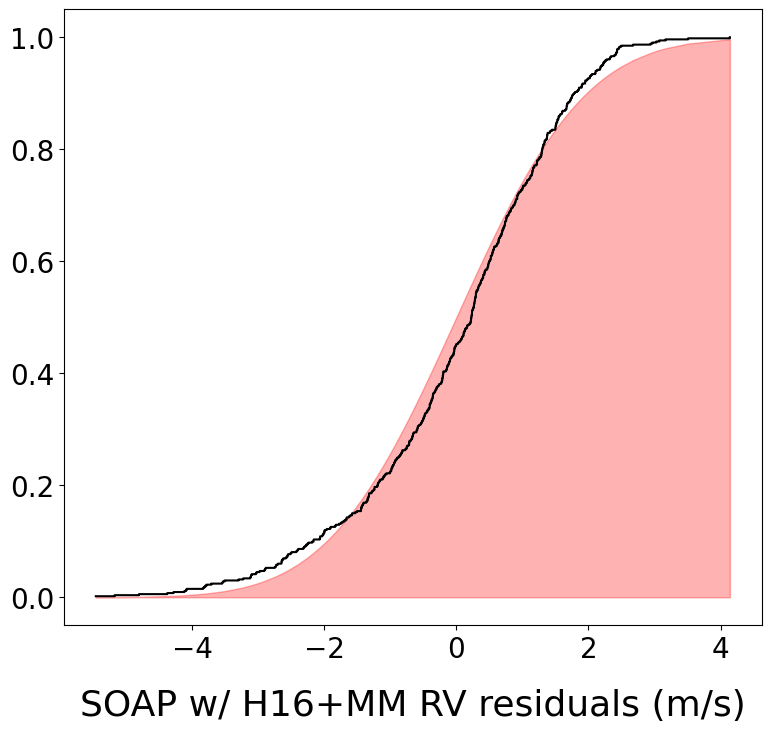}
     \end{subfigure}

   \end{minipage}

\caption{Left: Empirical CDF (black curve) of the RV residuals for the \texttt{SOAPv4} simulation based on MM active-region identification. Middle: Same for the H16 identification. Right: Same for the combined H16+MM approach. In all panels, the shaded red area shows the cumulative distribution of the best-fitting normal distribution.}
\label{fig:cumulative_residuals}
\end{figure*}

For the MM-based simulations, we obtain $D$ statistic $= 0.025$ and $p$-value $= 0.88$, indicating strong consistency with a Gaussian distribution. The H16 simulations yield $D$ statistic $= 0.052$ and $p$-value $= 0.098$, which does not allow us to rule out the null hypothesis at the 5 \% significance level and therefore suggests approximate normality. In contrast, the H16+MM case shows a larger deviation ($D$ statistic $= 0.066$, $p$-value $= 0.018$), for which the null hypothesis is rejected, indicating a statistically significant departure from Gaussian residuals. These results are presented in Table~\ref{tab:ks_results} and the CDFs are shown in Fig.~\ref{fig:cumulative_residuals}.

\begin{table}[h!]
\centering\renewcommand{\arraystretch}{1.3}
\caption{KS test results for the normality of the RV residuals.}

\begin{tabular}{lcc}

\hline
Simulation & $D$ statistic & $p$-value \\
\hline
MM        & 0.025 & 0.88  \\
H16       & 0.052 & 0.098 \\
H16+MM    & 0.066 & 0.018 \\
\hline
\end{tabular}
\label{tab:ks_results}

\end{table}

\section{RV comparison with \texttt{ARVE} and \texttt{YARARA}
}
\label{append}
 We present a comparison of the simulated RVs and those extracted with \texttt{ARVE} and \texttt{YARARA}. The std of the residuals shows a slight increase compared to that obtained using the DRS RVs across all simulation scenarios, reaching $1.00$ m/s for \texttt{ARVE} and $1.05$ m/s for \texttt{YARARA} in the case of simulations considering the MM method, $1.61$ m/s for \texttt{ARVE} and $1.69$ m/s for \texttt{YARARA} for the H16-based simulations, and $1.77$ m/s for \texttt{ARVE} and $1.85$ m/s for \texttt{YARARA} for the simulations combining H16 and MM. This discrepancy likely arises from differences in how RVs are computed among the pipelines, particularly since both \texttt{ARVE} and \texttt{YARARA} derive RVs from spectra that have been corrected (see Sect~\ref{data}).
\begin{table}
  \centering\renewcommand{\arraystretch}{1.3}
\caption{Standard deviation of RV residuals (m/s) between the \texttt{SOAPv4} simulations considering the different identification methods and the different extraction pipelines.}
\begin{tabular}{lccc}
\hline
Identification method & DRS & ARVE & YARARA \\
\hline
MM (AIA \& HMI) & 0.91 & 1.00 & 1.05 \\
H16 & 1.40 & 1.61 & 1.69 \\
H16 + MM & 1.54 & 1.77 & 1.85 \\
\hline
\end{tabular}
\label{tab:RV_std}
\end{table}

\begin{figure*}
\centering

    \centering
    \includegraphics[width=\textwidth]{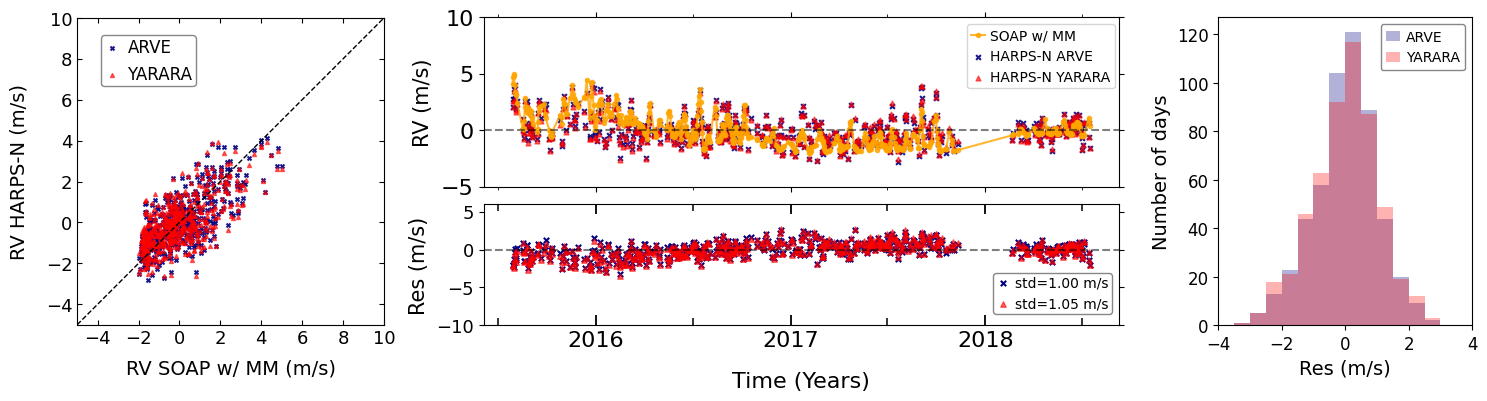}

    \caption{Left: Comparison of RVs from HARPS-N extracted with \texttt{ARVE} (blue) and with \texttt{YARARA} (red), and RV \texttt{SOAPv4} simulations based on the MM method both for spots and plages (\texttt{SOAP} with MM). Middle: RV time series from HARPS-N \texttt{ARVE} (blue) and \texttt{YARARA} (red) and from \texttt{SOAP} with MM (yellow; top), together with the residuals, calculated as the difference between observed and simulated RVs and their std value (bottom). Right: Histogram of the residuals between \texttt{SOAPv4} RVs and \texttt{ARVE} (blue) and between \texttt{SOAPv4} RVs and \texttt{YARARA} (red).}
\label{arveMM}
\end{figure*}
\begin{figure*}
\centering
    \includegraphics[width=\textwidth]{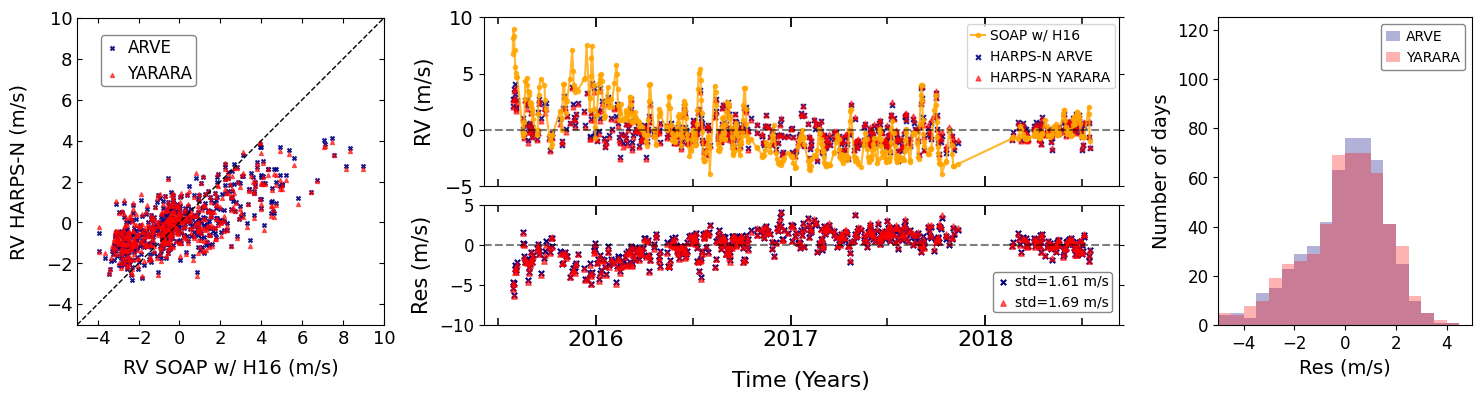}
    \caption{Same as Fig.~\ref{arveMM} but for \texttt{SOAPv4} simulations based on the H16 method.}
\label{arveh16}

\end{figure*}
\begin{figure*}
\centering

    \centering
    \includegraphics[width=\textwidth]{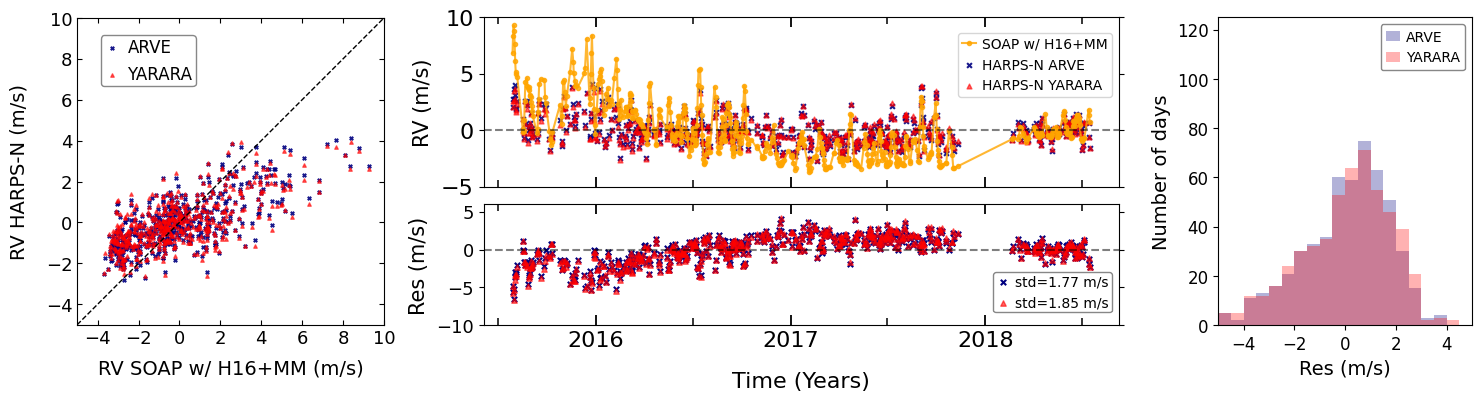}

    \caption{Same as Fig.~\ref{arveMM} but for \texttt{SOAPv4} simulations based on H16+MM.}
\label{arveh16MM}

\end{figure*}

\end{appendix}

\end{document}